 \documentclass[final,3p,times,twocolumn]{elsarticle}
 
\usepackage{lineno,hyperref}
\modulolinenumbers[5]
\hypersetup{
  colorlinks=true, % Enable colored links
  linkcolor=blue,  % Color for normal internal links
  citecolor=green, % Color for bibliographic citations
  urlcolor=magenta, % Color for URLs
}
\journal{}
\usepackage{graphicx}
\usepackage{nicefrac}
 \usepackage{mathtools}
 \usepackage{multicol}
\usepackage{lipsum}
\usepackage{color}
\usepackage{color,soul}
\usepackage{cuted}
\usepackage{amssymb}
\usepackage{amsmath}
\usepackage{float}
\begin{document}
\begin{frontmatter}
\title{Influence of Refractive Index Distribution on Multimode Soliton Dynamics and Condensation in GRIN-MMFs}
\author{Love Kumar Sharma}
\author{Vishwa Pal}
\ead{vishwa.pal@iitrpr.ac.in}
\address{Laser Research Lab, Department of Physics, Indian Institute of Technology Ropar, Rupnagar 140001, Punjab, India}

\begin{abstract}
Optical solitons propagating through a multimode fiber represents one of the most fascinating class of objects exhibiting peculiar properties, with widespread potential for applications. We theoretically investigate the effect of the core refractive index distribution, characterized by the index exponent $\alpha$, on the evolution of multimode (MM) soliton beams and their peculiar properties in graded-index multimode fibers. Our analysis reveals an optimal range $\alpha = 2.04$–$2.08$, within which MM solitons with minimum pulsewidth and characteristic energy are formed, owing to reduced modal walk-off and enhanced intermodal nonlinear interactions. Within this regime, the MM soliton undergoes efficient spatial condensation into the fundamental mode, resulting in a well-defined quasi-Gaussian output intensity profile.  Notably, for some particular values of $\alpha$, we observe a reversal of conventional energy flow associated with MM soliton condensation, leading to the transfer of energy toward higher-order modes, akin to the thermalization of MM optical fields into negative-temperature equilibrium states. Furthermore, we show that the characteristic Raman-induced spectral redshift of MM solitons can be controlled by tailoring the refractive index distribution. Our results highlight the refractive index distribution as a key control parameter governing MM soliton dynamics and their condensation behavior and are expected to be relevant for the design and optimization of MM fiber-based systems where controlled spatiotemporal dynamics are desired.

\end{abstract}
%%%%%%%%%%%%%%%%%%%%%%%%%%%%%%%%%%%%%%%%%%%%%%%%%%%%%%%%%%%%%%%%%%%%%%%%%%%%%%%%%%%%%%%%%%%%%%%%%%%%%%%%%%%%%%%%%%%

\end{frontmatter}

%%%%%%%%%%%%%%%%%%%%%%%%%%%%%%%%%%%%%%%%%%%%%%%%%%%%%%%%%%%%%%%%%%%%%%%%%%%%%%%%%%%%%%%,%%%%%%%%%%%%%%%%%%%%%%%%%
\section{Introduction}
\label{Intro}
\noindent
Recent years have witnessed a renewed research interest in the study of complex spatiotemporal dynamics arising from nonlinear pulse propagation in multimode fibers (MMFs). Unlike single-mode fibers (SMFs), MMFs provide additional degrees of freedom for light control and manipulation, enabling a wide range of emerging applications \cite{krupa2019multimode}. However, these additional dimensions also introduces significant complexities, making the control of nonlinear inter-modal interactions increasingly challenging as the number of supported mode increases \cite{shen2023roadmap}. In this context, sustained efforts have been directed toward understanding and harnessing nonlinear intermodal interactions for applications ranging from establishing new platforms for high-power laser sources to the development of space-division multiplexed optical communication technologies \cite{li2014space, cao2023spatiotemporal}. As a result, MMFs have emerged as a rich and versatile platform for the investigation of various novel nonlinear phenomena with no direct counterparts in traditional single-mode systems, including phenomena such as geometric parametric instabilities \cite{krupa2016observation}, intermodal four-wave mixing \cite{dupiol2017far}, Kerr beam self-cleaning \cite{krupa2017spatial, sharma2024kerr}, spatiotemporal mode-locking \cite{wright2017spatiotemporal}, multimode solitons \cite{sun2024multimode}, and novel class of dispersive Cherenkov radiation \cite{eftekhar2021general}, to mention a few.

Solitons are nonlinear wave packets that exhibit particle-like behavior and complex nonlinear dynamics, and are ubiquitous across a wide range of physical systems, including fluids \cite{osborne1980internal}, plasmas \cite{zabusky1965interaction}, Bose-Einstein condensates \cite{denschlag2000generating}, nonlinear lattices \cite{kartashov2011solitons}, and optical fibers \cite{hasegawa1989optical}. In SMFs, temporal solitons arise from a balance between Kerr nonlinearity and chromatic dispersion of the fiber medium. For ultrashort pulses, the soliton spectrum broadens significantly, causing the longer wavelength spectral components to experience Raman-induced gain at the expense of the shorter-wavelength components \cite{mitschke1986discovery}. This leads to a continuous redshift of the soliton spectrum, known as soliton self-frequency shift (SSFS), which has been extensively studied, and is demonstrated to have applications such as in development of wavelength-tunable femtosecond laser sources, tunable optical delay lines, etc \cite{lee2008soliton}.

In contrast, MMFs supports MM solitons which are synchronized, nondispersive pulses spread across multiple spatial modes of a MMF \cite{renninger2013optical}. Previous studies have investigated their generation, propagation and spatiotemporal dynamics, showing that MM solitons tend to spatially localize itself into a fundamental mode soliton due to Raman-induced energy transfer \cite{wright2015spatiotemporal}. Moreover, the interplay of strong nonlinear effects during ultrashort pulse propagation in MMFs enables the generation of a broadband, tunable, multi-octave supercontinuum spectra \cite{wright2015controllable}. Manipulating MM soliton dynamics and their properties remains a key challenge for many applications involving MM solitons and solitary waves. In this context, recent studies have shown that MM solitons in GRIN-MMFs exhibit peculiar properties in terms of their pulsewidths and characteristic energies \cite{zitelli2021conditions}. Notably, the MM soliton pulsewidth at the output of a MMF is found to be independent of the input pulse duration, and is dependent only on the wavelength of the input pulse, and the input coupling conditions \cite{zitelli2024optical, sharma2025impact}. Furthermore, it has also been demonstrated recently that MM solitons undergo irreversible spatial condensation into a fundamental mode soliton, by virtue of the combined effects of Raman and Kerr nonlinearities, which facilitate gradual energy transfer towards the fundamental mode \cite{zitelli2021single}. The refractive index distribution within the core of a GRIN-MMF plays a critical role in determining modal confinement and the spatial overlap among guided modes \cite{dai2024influence}, thereby governing the efficiency of nonlinear intermodal interactions and ultimately dictating the formation, propagation, and spatiotemporal characteristics of MM solitons. However, a comprehensive understanding of the nature of dependence of MM soliton dynamics and their peculiar properties on the refractive index distribution within the fiber core remains unexplored. Addressing this gap forms the primary objective of the present work.

In this work, we theoretically investigate how the peculiar properties of MM solitons and their condensation into the fundamental mode soliton depend on the refractive index distribution within the core region of a GRIN-MMF. We identify an optimal parameter range $(\alpha=2.04-2.08)$, within which reduced modal walk-off enables strong intermodal interactions, leading to the formation of MM solitons at relatively lower values of input energy. In this regime, the MM soliton undergoes efficient spatial condensation into the fundamental mode Raman soliton through gradual energy accumulation. In contrast, for $\alpha = 1.80$, no such spatial condensation is observed. Instead, a gradual net transfer of energy from lower-order modes (LOMs) to higher-order modes (HOMs) is found, indicating a reversal of the conventional energy flow associated with MM soliton condensation. This demonstrates that the refractive index exponent $(\alpha)$ governs the strength of nonlinear intermodal interactions and dictates the direction of energy transfer. To the best of our knowledge, this is the first report of such reverse energy flow in the MM soliton regime, with similar behavior previously reported only in the normal dispersion regime where MM solitons do not form and intermodal four-wave mixing (FWM) dominates Kerr beam self-cleaning, and has been interpreted as thermalization of classical nonlinear waves towards a negative temperature equilibrium state \cite{baudin2023observation}. In addition, we demonstrate that the characteristic Raman-induced spectral redshift of MM solitons can be effectively controlled by tailoring the refractive index distribution. Overall, our results highlight the refractive index distribution as a key parameter for controlling MM soliton dynamics and their condensation behavior.

The paper is organized as follows. Section 2 introduces the theoretical model and physical parameters used to describe nonlinear pulse propagation in MMFs. Section 3 presents a detailed numerical investigation of MM soliton dynamics as a function of the refractive index distribution within the fiber core. In particular, Section 3.1 discusses the spatiotemporal dynamics of MM soliton formation, Section 3.2 examines the influence of the refractive index distribution on MM soliton properties such as pulsewidth and spectral evolution, and Section 3.3 analyzes the impact of the refractive index distribution on the spatial evolution of MM soliton beams and the associated energy flow dynamics. Finally, Section 4 summarizes the main conclusions of this manuscript.

\section{Theoretical modeling and parameters}
\label{theo_desc}
\noindent
An optical field propagating through a MMF supporting M spatial modes can be expressed in terms of its modal basis as \cite{mafi2012pulse}:
\begin{equation}
E(x,y,z,t)=\sum_{p=1}^{M}F_{p}(x,y) \hspace{2pt}A_{p}(z,t),
~~~~~\label{eq1}
\end{equation}
where $F_{p}(x,y)$ represents the transverse spatial field distribution of the $p^{\mathrm{th}}$ mode, and $A_{p}(z,t)$ denotes its corresponding slowly varying field envelope. 

A numerical model based on such coupled-mode formalism is well suited for studying the spatiotemporal dynamics of nonlinear pulse propagation in MMFs, particularly when a limited number of modes are involved. This approach involves solving a set of N-coupled partial differential equations, each governing the evolution of a modal field envelope $A_{p}$. Such a framework provides valuable insights into the mode coupling dynamics, energy redistribution, and intermodal energy transfer, helping us to understand the modal dynamics associated with many intriguing nonlinear phenomena in MMFs. In this work, we employ a model based on the generalized multimode nonlinear Schrodinger equations (GMMNLSEs) to investigate the influence of refractive index distribution on the MM soliton dynamics and their spatial condensation. Under the slowly varying envelope and paraxial approximations, the evolution of the field envelope $A_{p}$ of a given mode $p$ is governed by the interplay of linear dispersive and nonlinear effects as \cite{poletti2008description,wright2017multimode}:
\begin{equation}
\frac{\partial A_p(z,t)}{\partial z}
= \mathcal{D}\{A_p(z,t)\} + \mathcal{N}\{A_p(z,t)\},
\label{eq2}
\end{equation}
\noindent
where the dispersion operator is given by:
\begin{equation}
\mathcal{D}\{A_p\} =
i\,\delta\beta^{(p)}_{0}\,A_{p}
- \delta\beta^{(p)}_{1}\,\partial_{t} A_{p}
+ \sum_{m=2}^{3} i^{\,m+1}\,\frac{\beta^{(p)}_{m}}{m!}\,\partial_t^{m} A_{p}.
\label{eq3}
\end{equation}
Here, the first two terms represent, respectively, the difference in propagation constants and inverse group velocities of the $p^{\mathrm{th}}$ mode with respect to the fundamental mode. Together, they account for phase mismatch and group-velocity mismatch, the latter giving rise to temporal walk-off and a progressive temporal separation of modal components during propagation. The third term incorporates higher-order dispersion effects within each mode.

\noindent
Similarly, the nonlinear operator is expressed as:
\begin{multline}
\mathcal{N}\{A_p\} =
i\frac{n_{2}\omega_{0}}{c}
\Big(1+\frac{i}{\omega_{0}}\partial_{t}\Big)
\sum_{l,m,n}^{N}
\Big[
(1-f_{R})S^{K}_{plmn}A_{l}A_{m}A^{*}_{n}
\\
+ f_{R}S^{R}_{plmn}A_{l}
\int_{-\infty}^{t} d\tau\, h_{R}(\tau)\,
A_{m}(z,t-\tau)A^{*}_{n}(z,t-\tau)
\Big].
\label{eq4}
\end{multline}
Here, $f_{R}$ denotes the fractional Raman contribution to the total nonlinearity for silica-based GRIN-MMFs, $\omega_{0}$ is the central angular frequency of the input pulse, and $n_{2}=2.37\times 10^{-20} \thinspace {{\mathrm{m^2}}/{\mathrm{W}}}$ is the nonlinear index of refraction. The delayed, convoluted Raman response function of the fiber medium is given by \cite{stolen1989raman}:
\begin{equation}
h_R(t)
= \frac{\tau_1^2 + \tau_2^2}{\tau_1 \tau_2^2}
\, e^{-t/\tau_2}
\sin\!\left(\frac{t}{\tau_1}\right),
\quad t \ge 0,
\label{eq5}
\end{equation}
where $\tau_1 = 12.2~\mathrm{fs}$ and $\tau_2 = 32~\mathrm{fs}$ are the characteristic parameters of silica.

The coefficients $S^{\mathrm{K}}_{plmn}$ and $S^{\mathrm{R}}_{plmn}$, respectively, represents the nonlinear modal overlap integrals that govern Kerr and Raman nonlinearity of the medium among spatial modes $p$, $l$, $m$, and $n$, and are defined as:
\begin{equation}
S_{plmn}
=
\frac{
\displaystyle \int \mathrm{d}x\,\mathrm{d}y\,
F_p F_l F_m F_n
}{
\sqrt{
\displaystyle
\int \mathrm{d}x\,\mathrm{d}y\, F_p^2
\int \mathrm{d}x\,\mathrm{d}y\, F_l^2
\int \mathrm{d}x\,\mathrm{d}y\, F_m^2
\int \mathrm{d}x\,\mathrm{d}y\, F_n^2
}
}.
\label{eq6}
\end{equation}

Direct numerical integration of Eq.\,(\ref{eq2}) using conventional split-step Fourier methods is computationally very demanding and becomes inefficient for MM systems even when a fewer number of modes are involved. To address this, we employ an optimized massively parallel algorithm (MPA), which significantly accelerates the iterative process by distributing the computational load across mutliple processors \cite{kazakov2023parallelization}. For the simulations presented in this work, we numerically model light propagation through a $150$\thinspace{m} long GRIN-MMF with a core diameter of $62.5$\thinspace$\mu$m and a numerical aperture of $0.2$. The choice of a relatively long propagation distance ensures the complete development of nonlinear spatiotemporal dynamics, including MM soliton formation and its condensation, while also poses significant computational demands. The dispersion coefficients up to third order are calculated by modeling the fiber as a germanium-doped silica medium. A longitudinal step size of $50\thinspace\mu$m, a temporal resolution of $3.9$\thinspace{fs} with a time window of $40$\thinspace{ps} are used to ensure accurate resolution of ultrashort pulse dynamics. We have considered both Raman nonlinearity and self-steepening effects in the simulations. 

It is important to note that Eq.\,(\ref{eq2}) inherently couples the propagating modal field envelopes through nonlinear interactions. The Raman contribution of such nonlinear interaction is governed by the coupling coefficients $S^{\mathrm{R}}_{plmn}$, whereas Kerr nonlinear interactions are described by $S^{\mathrm{K}}_{plmn}$. In particular, self-phase modulation (SPM) corresponds to terms of the form $S^{\mathrm{K}}_{pppp}$, with effective modal area defined as $A_\mathrm{eff}=1/S^{\mathrm{K}}_{pppp}$ for that particular mode. Cross-phase modulation (XPM) arises from terms such as ${S^{\mathrm{K}}_{ppnn}}$ and ${S^{\mathrm{K}}_{pnpn}}$ for $n\neq p$, while all remaining nonlinear coupling terms correspond to intermodal four-wave mixing (FWM) processes, which can cause transfer of energy among the co-propagating modes. 

The fiber modes considered in this study are linearly polarized (LP), obtained under the scalar field approximation. This allows the co-propagating modal fields to be represented by real-valued functions $F_{p}(x,y)$ with corresponding propagation constants $\beta_{p}$. Within this framework, the nonlinear coupling coefficients are real, non-negative, and independent of polarization, simplifying the analysis while retaining the essential physics of MM interactions.

\begin{figure}[h!]
 \includegraphics[width=7.8 cm]{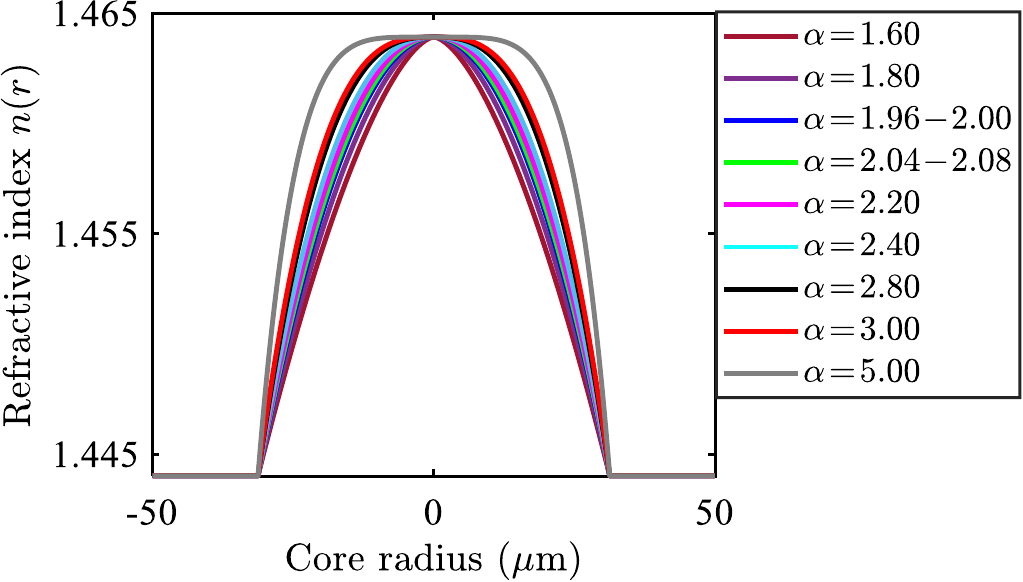}
\centering
\caption{Refractive index distribution within the core region of a GRIN-MMF for different values of the index exponent ($\alpha$), illustrating the transition from near-parabolic to increasingly flattened or peaked profiles as $\alpha$ is varied.}
\label{fig1}
\end{figure}

The refractive index distribution within the core of the GRIN-MMF is modeled using a generalized profile of the form:
\begin{equation}
n^{2}(r) = n_{\mathrm{co}}^{2} \left[1 - 2\Delta \left(\frac{r}{a}\right)^{\alpha} \right],
\end{equation}
where $r=\sqrt{x^{2}+y^{2}}$ is the radial coordinate, $a$ is the core radius of the fiber, and $\Delta = (n_{\mathrm{co}} - n_{\mathrm{cl}})/n_{\mathrm{co}}$ represents the relative refractive index difference between the core and the cladding, with $n_{\mathrm{co}}$ and $n_{\mathrm{cl}}$ being the refractive indices of the core center and cladding, respectively. The exponent $\alpha$ characterizes the shape of the refractive index distribution, with $\alpha = 2.00$ corresponding to the ideal parabolic distribution. In this work, $\alpha$ is varied over the range $\alpha = 1.60$--$3.00$ to systematically investigate deviations from the parabolic distribution and their influence on modal confinement, intermodal interactions, and MM soliton dynamics. Figure \ref{fig1} illustrates the corresponding refractive index profiles for different values of $\alpha$.

\noindent
A Gaussian input pulse centered at $1550$\thinspace{nm} is launched into the considered GRIN-MMF, exciting a total of $10$ spatial modes. The input coupling conditions are chosen such that $80$\thinspace{\%} of input energy is distributed among the first six modes, while the remaining $20$\thinspace{\%} is allocated to the remaining four modes. This choice of a limited number of spatial modes strikes a balance between capturing the essential physics of intermodal nonlinear interactions and maintaining computational efficiency, while still enabling accurate investigation of MM soliton formation and energy redistribution dynamics.
\section{Results and discussions} \label{results}
\subsection{Spatiotemporal dynamics of MM soliton formation}
\begin{figure*}[h!]
\centering
\includegraphics[width = 17 cm, keepaspectratio = true]{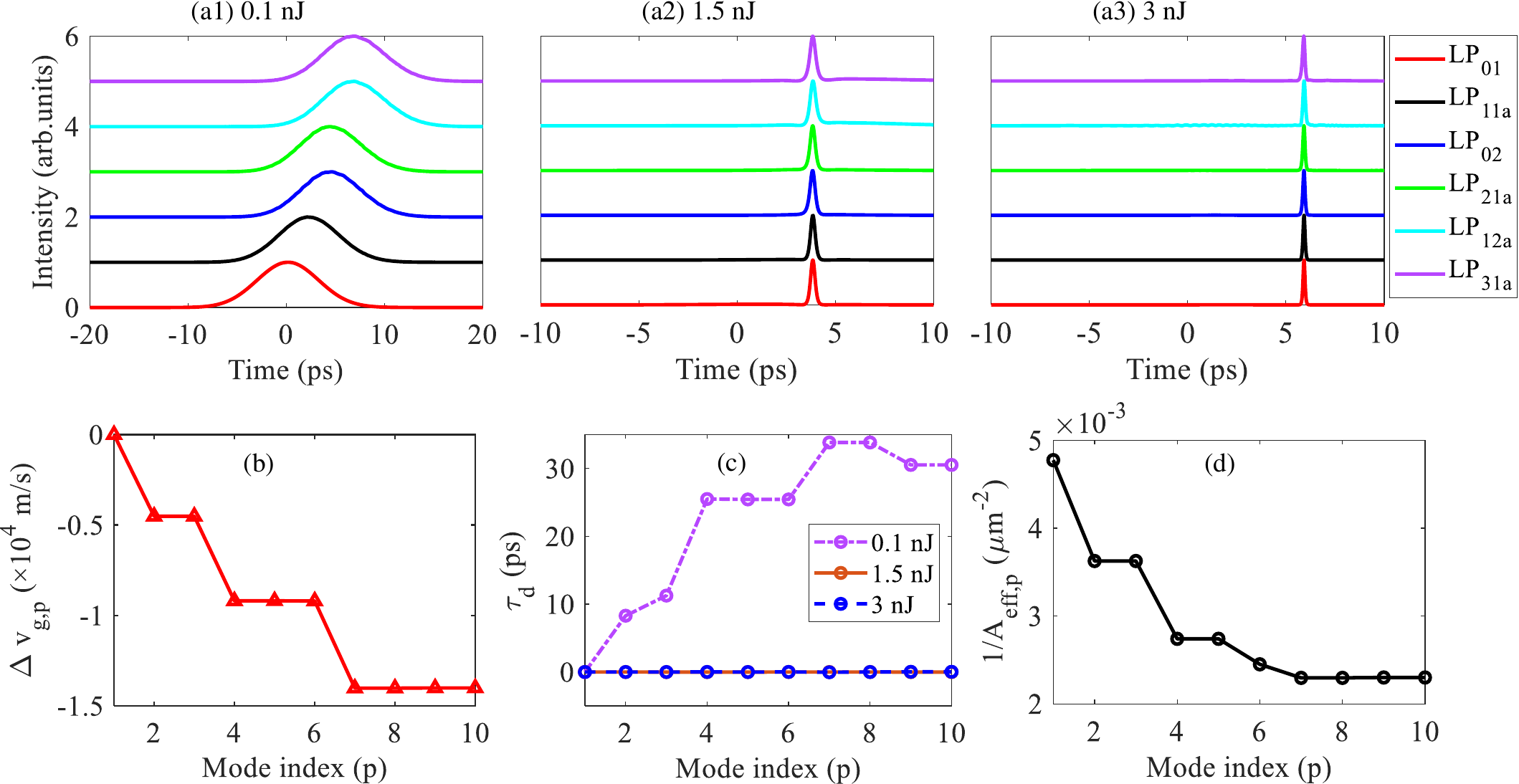}
\caption {Evolution of the mode-resolved temporal intensity profiles of the excited spatial modes in the considered GRIN-MMF with increasing input energy: (a1) linear propagation regime ($0.1$\thinspace{nJ}) after $20$\thinspace{m}, where the modes broaden temporally due to chromatic dispersion and separate owing to intermodal dispersion; (a2) intermediate quasi-MM soliton regime ($1.5$\thinspace{nJ}) after $20$\thinspace{m}, where temporal compression occurs due to IM-FWM along with the onset of spectral and temporal synchronization; and (a3) MM soliton regime ($3$\thinspace{nJ}) after $30$\thinspace{m}, where the modes become temporally locked and propagate with a common group velocity. The temporal intensity profiles are normalized to the peak intensity of each mode and are vertically offset for clarity. (b) Relative group velocity (with respect to the fundamental mode) as a function of mode index, showing the nearly uniform spacing of modal group velocities. The absolute group velocity of the fundamental mode is approximately $2 \times 10^{8}\,\mathrm{m/s}$, whereas the relative group-velocity differences between modes are of the order of $10^{4}\,\mathrm{m/s}$; (c) relative temporal delay $\tau_d$ as a function of mode index after $120$\thinspace{m} of propagation, illustrating the increase in intermodal delay (temporal walk-off) at low input energies, where the modes progressively separate, and the transition to perfect temporal synchronization at higher input energies; and (d) inverse effective mode area $1/A_{\mathrm{eff},p}$ as a function of mode index, highlighting the mode-dependent nonlinear interaction strength.}
\label{fig2}
\end{figure*}

Figures \ref{fig2}(a1)--\ref{fig2}(a3) illustrates the evolution of the mode-resolved temporal dynamics with increasing input energy across three distinct propagation regimes for $\alpha=2.08$. In the linear propagation regime (with input energy $E_{\mathrm{in}} < 1$\thinspace{nJ}), the individual modal pulse undergo temporal broadening due to chromatic dispersion and separate from each other because of intermodal dispersion. In this regime, Kerr nonlinearity is insignificant, and no appreciable intermodal energy exchange is observed. This behavior is further quantified with the help of Fig. \ref{fig2}(b), where the relative group velocity of a particular mode with respect to the fundamental mode is shown. The nearly uniform variation of the relative group velocity with mode index indicates that the modal group velocities are approximately equally spaced, with HOMs propagating at lower group velocities compared to the fundamental mode, leading to the observed temporal walk-off. The case of $0.1$\thinspace{nJ} in Fig. \ref{fig2}(c) further highlights the relative temporal delay accumulated by individual modes after propagation through $120$\thinspace{m} of the fiber, consistent with their relative group velocities.

Figure \ref{fig2}(d) shows the variation of the inverse effective mode area, $1/A_{\mathrm{eff},p}$, as a function of mode index, which serves as a direct measure of the nonlinear interaction strength of each mode. It is evident that $1/A_{\mathrm{eff},p}$ decreases with increasing mode index, indicating that LOMs possess stronger nonlinear interaction compared to HOMs. This behavior arises from the larger spatial extent of HOMs, which results in increased effective mode area and consequently weaker nonlinearity. Since the nonlinear coefficient $\gamma_p \propto 1/A_{\mathrm{eff},p}$, this implies that nonlinear processes such as IM-FWM and Raman scattering are more efficient in LOMs.

As the input energy increases, nonlinear effects become significant, leading to the partial compensation of the dispersion-induced pulse broadening. This gives rise to an intermediate quasi-MM soliton regime (Fig. \ref{fig2}(a2)), where the pulses begin to compress temporally and reach pulsewidth values roughly two or three times of those typically observed in a GRIN-MMF. In this regime, IM-FWM facilitates energy transfer predominantly from HOMs to LOMs, resulting in enhanced modal synchronization. This trend is also reflected in Fig. \ref{fig2}(c) with case of $1.5$\thinspace{nJ}, where the relative temporal delay between modes exhibits near-perfect temporal synchronization and suppression of intermodal walk-off.

As the input energy increases further, the system enters the MM soliton regime (Fig. \ref{fig2}(a3)), where a MM soliton consisting of many modes emerges. The MM soliton is characterized by non-dispersive, synchronized pulses spread across various spatial modes of a MMF. In this regime, pulses across different spatial modes become temporally locked and propagate together with a common group velocity. This behavior is also clearly evidenced in Fig. \ref{fig2}(c) with case of $3$\thinspace{nJ}, where the relative temporal delay across modes approaches zero, confirming perfect temporal synchronization between the modes. 

\begin{figure}[htbp]
\centering
\includegraphics[width=7.8 cm]{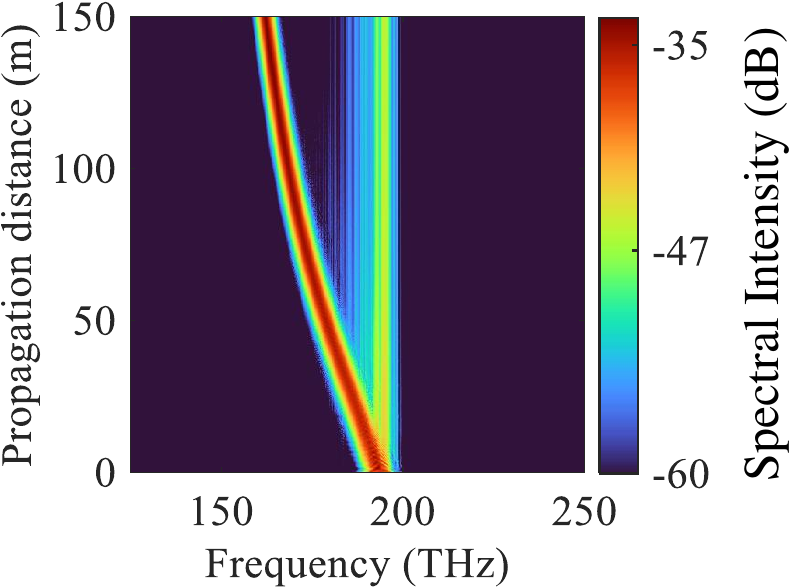}
\caption{Spectral evolution of a $200$\thinspace{fs}, $3.5$\thinspace{nJ} Gaussian input pulse centered at $193.14$\thinspace{THz} during propagation through a $150$\thinspace{m} long GRIN-MMF. A MM soliton comprising of many modes form after approximately $45$\thinspace{cm} of propagation. A small portion of the input energy is emitted as dispersive radiation near the pump frequency. Upon further propagation, the MM soliton undergoes a continuous Raman-induced redshift due to SSFS.}
\label{fig3}
\end{figure}

Figure \ref{fig3} shows the spectral evolution of a Gaussian input pulse of initial temporal width $200$\thinspace{fs}, central frequency $193.14$\thinspace{THz}, and energy $3.5$\thinspace{nJ} during propagation through a $150$\thinspace{m} long GRIN-MMF. As seen from the figure, a MM soliton involving many modes is formed after approximately $45$\thinspace{cm} of propagation. During this initial stage, a small fraction of the input energy is emitted as dispersive wave radiation, which remains localized near the pump frequency and propagates with distinct group velocities. Following its formation, the MM soliton undergoes a continuous Raman-induced spectral redshift as it propagates along the fiber. This behavior is characteristic of the well-known SSFS mechanism \cite{mitschke1986discovery}, arising from intra-pulse Raman scattering, and results in a progressive shift of the soliton spectrum toward longer wavelengths.

The dynamic interplay of IM-FWM and IM-Raman scattering drives a continuous redistribution of energy, leading to its gradual accumulation in the fundamental mode. This process ultimately results in the formation of a spatially condensed, stable MM soliton. Such an abrupt condensation of a MM soliton beam into a fundamental Raman soliton has recently been interpreted within a statistical mechanics framework, where the MM field is modeled as an ensemble of indistinguishable energy packets distributed among degenerate modal groups \cite{zitelli2024statistics}. The corresponding equilibrium modal distribution is given by \cite{zitelli2025new}:

\begin{equation}
|f_i|^2
=
\frac{2(g_i - 1)}{g_i \gamma}
\frac{1}{\exp\!\left(-\frac{\mu + \epsilon_i}{T}\right) - 1}.
\label{eq7}
\end{equation}
Here, $|f_i|^2$ denotes the power fraction carried by the $i^{\mathrm{th}}$ degenerate modal group, while $g_i$ is its degeneracy, i.e., the number of modes sharing the same propagation constant. The quantity $\epsilon_i$ represents the corresponding modal eigenvalue (or propagation constant offset) of the group. The parameters $\mu$ and $T$ denote the effective chemical potential and optical temperature of the system, respectively, arising from the entropy maximization procedure. The factor $\gamma$ is a normalization constant ensuring conservation of the total optical power across all modes.

It has been also recently shown that MM solitons exhibit peculiar properties in terms of their pulsewdith and characteristic energy. In particular, the output pulsewidth and energy of a MM soliton in a GRIN-MMF is independent of the input pulse duration and is only governed by the dispersive properties of the fiber \cite{zitelli2022characterization,sharma2025impact}.

\subsection{Effect of refractive index distribution on the properties of MM solitons}
\noindent
The refractive index distribution within the core region of a GRIN-MMF plays a crucial role in governing the evolution of MM soliton dynamics and their spatiotemporal characteristics. In this section, we numerically investigate the dependence of MM soliton pulsewidth and soliton wavelength on the core refractive index distribution by varying the index exponent $(\alpha)$ over the range $\alpha=1.60-3.00$. The corresponding variations in the refractive index distribution is shown in Fig. \ref{fig1}.

\begin{figure*}[h!]
 \includegraphics[width=17 cm]{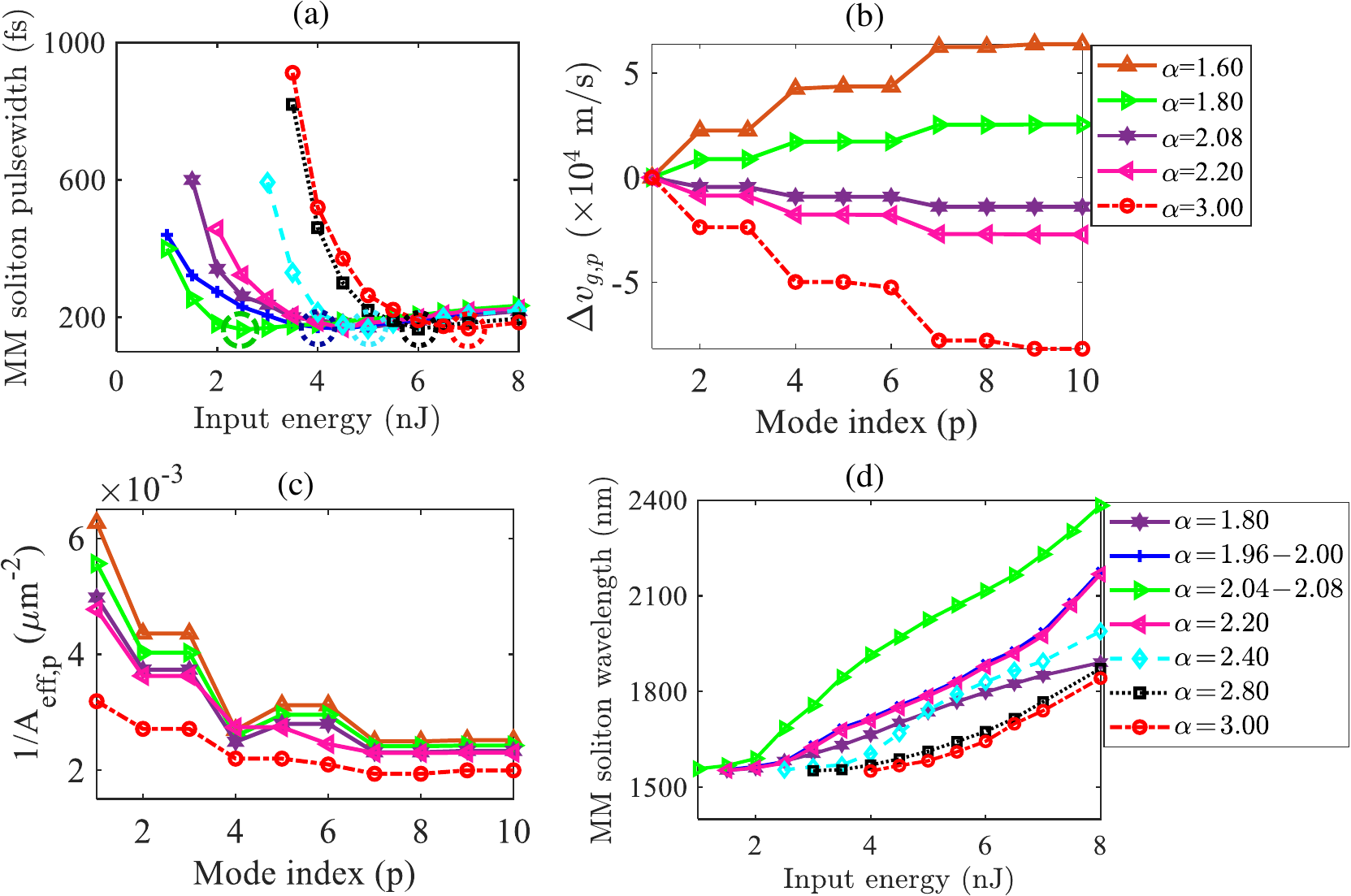}
\centering
\caption{Evolution of modal and MM soliton characteristics in a GRIN-MMF for different refractive index profiles characterized by the index exponent $\alpha$. (a) MM soliton pulsewidth as a function of input energy at the output of a $150$\thinspace{m} long fiber, showing the existence of a minimum pulsewidth corresponding to MM soliton formation. The circled markers indicate the input energies at which the minimum pulsewidth is achieved for each value of $\alpha$. (b) Relative group-velocity mismatch $\Delta v_{g,p}$ as a function of mode index, representing the intrinsic modal walk-off determined by the refractive index distribution. (c) Inverse effective modal area $1/A_{\mathrm{eff},p}$ versus mode index, indicating the variation of nonlinear interaction strength across modes for different values of $\alpha$. (d) MM soliton central wavelength as a function of input energy at the fiber output, illustrating the Raman-induced spectral redshift with increasing energy. The lowest characteristic soliton energies and strongest spectral shifts are observed within the optimal range $\alpha \approx 2.04$--$2.08$, highlighting enhanced nonlinear coupling and reduced modal walk-off in this range.}
\label{fig4}
\end{figure*}
\noindent

The variation of MM soliton pulsewidth with input energy for different values of $\alpha$ is presented in Fig. \ref{fig4}(a). It is observed that MM soliton forms only above a threshold value of $\alpha$ and no MM soliton formation is observed for $\alpha \leq 1.60$. For higher values of $\alpha$ (up to $\alpha=3.00$), MM solitons are observed to form across all considered refractive index distributions; however, the characteristic MM soliton energy remains strongly dependent on $\alpha$. In particular, we identify an optimal parameter range for $\alpha=2.04-2.08$, within which MM solitons form at relatively lower input energies. 

To understand the underlying origin of this behavior, it is instructive to examine some of the modal characteristics governing the nonlinear dynamics of MM soliton formation, as shown in Fig. \ref{fig4}(b) and Fig. \ref{fig4}(c) . Figure \ref{fig4}(b) shows the variation of the relative group-velocity mismatch across different modes for various values of $\alpha$. It is observed that for values of $\alpha$ close to the optimal range ($\alpha=2.04-2.08$), the spread in group velocities among the modes is relatively small, indicating reduced temporal walk-off between modes. In contrast, for larger deviations from the parabolic profile (e.g., $\alpha=1.80$ or $\alpha=3.00$), the group-velocity mismatch increases significantly, leading to stronger temporal separation among modes. This increased walk-off weakens the effective intermodal interaction and may also inhibit efficient MM soliton formation.

Further insight is provided by Fig. \ref{fig4}(c), which shows the variation of the inverse effective modal area as a function of mode index. Since the nonlinear coefficient is inversely proportional to the effective modal area, higher values correspond to stronger intrinsic nonlinear interactions. It is observed that for LOMs, the inverse effective modal area increases with decreasing $\alpha$, indicating enhanced intrinsic nonlinearity due to stronger modal confinement. However, within the optimal $\alpha$ range, the inverse effective modal areas are relatively large and exhibit less variation across modes, resulting in strong and more balanced intermodal nonlinear interactions. In contrast, for non-optimal values of $\alpha$, the inverse effective modal areas become more uneven across modes, leading to an imbalance in the nonlinear interaction strengths.
For $\alpha \approx 1.60$, although the intrinsic nonlinear interaction strength remains relatively high for LOMs, the significantly increased modal walk-off reduces the temporal overlap among the modes. As a result, the effective nonlinear interaction length is substantially reduced, limiting intermodal energy exchange. Consequently, the intermodal nonlinear coupling becomes insufficient to sustain the formation of a stable MM soliton, explaining the absence of MM soliton formation in this regime.

Interestingly, the minimum output MM soliton pulsewidth is found to be approximately $165$\thinspace{fs}, and remains invariant across all refractive index distributions where MM soliton formation takes place, as shown in Fig. \ref{fig4}(a). This observation is consistent with earlier assertion that the MM soliton pulsewidth at the output of a GRIN-MMF is independent of the input pulsewidth, and is governed primarily by the dispersive properties of the fiber. Such an invariance of the MM soliton pulsewidth with respect to the input pulse duration has been previously explained using the so-called soliton walk-off theory \cite{zitelli2021conditions}. In a MMF, the spatiotemporal dynamics of MM solitons are governed by the interplay of three different characteristic length scales: the chromatic dispersion length, $L_{D} = \frac{T_{0}^{2}}{|\beta_{2}|}$, where $T_{0} = \frac{T_{\mathrm{FWHM}}}{1.763}$ is related to the input pulse duration; the mean modal walk-off length, $L_{W} = \frac{T_{0}}{\overline{\Delta \beta_{1}}(\lambda)}$, where $\overline{\Delta \beta_{1}}(\lambda)$ represents the power-weighted average group-velocity mismatch of the propagating modes with respect to the fundamental mode; and the nonlinear length, $L_{NL} = \frac{1}{\gamma P_{0}}$, where $\gamma$ is the effective nonlinear coefficient of the fiber and $P_{0}$ is the peak power of the input pulse. A stable MM soliton with a well-defined pulsewidth is formed when these three length scales become comparable, i.e., $L_{D} \sim L_{NL} \sim \mathrm{const} \cdot L_{W}$, where $\mathrm{const}$ is a proportionality factor that depends on the input coupling conditions \cite{zitelli2021conditions}.

At higher input energies, the MM soliton pulsewidth increases for all values of $\alpha$. For instance, the MM soliton pulsewidth rises from $165$\thinspace{fs} to $205$\thinspace{fs} as the input energy increases from $2.5$\thinspace{nJ} to $6$\thinspace{nJ} for $\alpha=2.08$. This behavior can be accredited to the continuous redshift of the MM soliton spectrum due to SSFS, wherein the central wavelength shifts from $1705$\thinspace{nm} to $2116$\thinspace{nm}. As the wavelength increases, the magnitude of the group-velocity dispersion (GVD) parameter also increases, leading to a corresponding increase in the MM soliton pulsewidth. For instance, the magnitude of $|\beta_{2}|$ for the fundamental mode increases from $28.14~\mathrm{ps^{2}/km}$ to $53.5~\mathrm{ps^{2}/km}$ as the wavelength shifts from $1550\thinspace{\mathrm{nm}}$ to $1750\thinspace{\mathrm{nm}}$, thereby resulting in the observed temporal broadening of the MM soliton.

\begin{figure*}[h!]
 \includegraphics[width=17 cm]{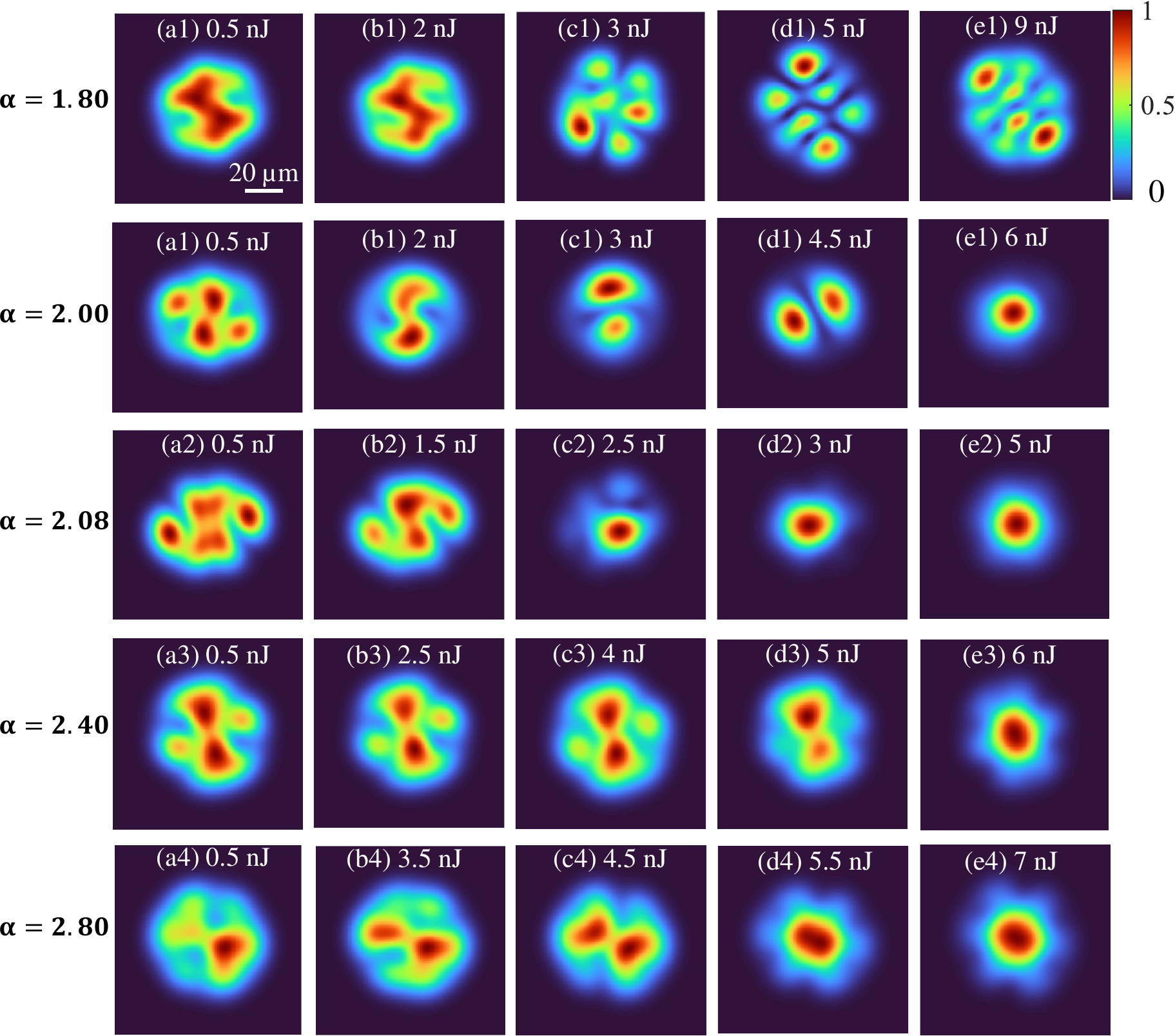}
\centering
\caption{Evolution of the spatial intensity profile of the MM beam at the output of a $150$\thinspace{m} long GRIN-MMF with increasing input energy for different refractive index distributions characterized by the index exponent $\alpha$. Each row corresponds to a fixed value of $\alpha$, while the columns represent increasing input energy. At low input energies, the output beam exhibits a speckled structure due to the incoherent superposition of multiple modes. As the input energy increases, nonlinear interactions lead to progressive spatial reorganization and eventual condensation into a quasi-Gaussian beam profile. Efficient condensation into the fundamental mode is observed for $\alpha \approx 2.04$--$2.08$, whereas for lower values of $\alpha$ (e.g., $\alpha = 1.80$), the beam remains MM and exhibits signatures of reverse energy flow toward HOMs.}
\label{fig5}
\end{figure*}
\noindent

We also investigate the evolution of the MM soliton wavelength with increasing input energy for different refractive index distributions, as shown in Fig. \ref{fig4}(d). It is observed that, at moderate value of input energies, Kerr nonlinearity induces spectral shifts in individual modes, enabling them to attain a common group velocity and become temporally synchronized, leading to the formation of a MM soliton. 

With further increase in input energy, the MM soliton undergoes a continuous Raman-induced spectral redshift due to SSFS. To quantify this behavior, we analyze the evolution of the MM soliton central wavelength as a function of input energy for different values of the index exponent $\alpha$. As shown in Fig. \ref{fig4}(d), MM solitons propagating in GRIN-MMFs with ($\alpha=2.04-2.08$) exhibit the strongest Raman-induced spectral redshift compared to other index profiles. For instance, the MM soliton central wavelength shifts from $1705$\thinspace{nm} at $2.5$\thinspace{nJ} to $2116$\thinspace{nm} at $6$\thinspace{nJ}. In contrast, for other values of $\alpha$, the redshift is significantly weaker over the same energy range and progressively decreases as $\alpha$ deviates from the optimal range, with the smallest spectral shift observed for $\alpha = 3.00$.

This variation in the extent of spectral redshift can be understood from the modal characteristics shown in Figs. \ref{fig4}(b) and \ref{fig4}(c). For values of $\alpha$ within the optimal range, the smaller relative group-velocity mismatch [Fig. \ref{fig4}(b)] ensures reduced modal walk-off and hence strong temporal synchronization among modes, enabling sustained nonlinear interaction over longer propagation distances. In addition, the inverse effective modal areas are relatively large and exhibit less variation across modes [Fig. \ref{fig4}(c)], indicating strong and more balanced intrinsic nonlinear interactions. These combined effects lead to enhanced Raman interaction and consequently a more pronounced SSFS.

In contrast, for non-optimal values of $\alpha$, the increased modal walk-off significantly reduces the temporal overlap among modes, thereby limiting the effective nonlinear interaction length. Although the intrinsic nonlinearity may remain comparable or even stronger for certain modes, the reduced interaction length weakens the overall intermodal coupling. As a result, the efficiency of Raman-induced spectral shifting is significantly diminished, leading to a comparatively weaker redshift.

This enhanced redshift also has a direct impact on the temporal characteristics of the MM soliton. As discussed earlier, the shift of the central wavelength toward longer wavelengths leads to an increase in GVD parameter, resulting in a corresponding broadening of the MM soliton pulsewidth at higher input energies.

\subsection{Effect of refractive index distribution on spatial condensation of MM solitons}
\begin{figure*}[h!].
 \includegraphics[width=16.5 cm]{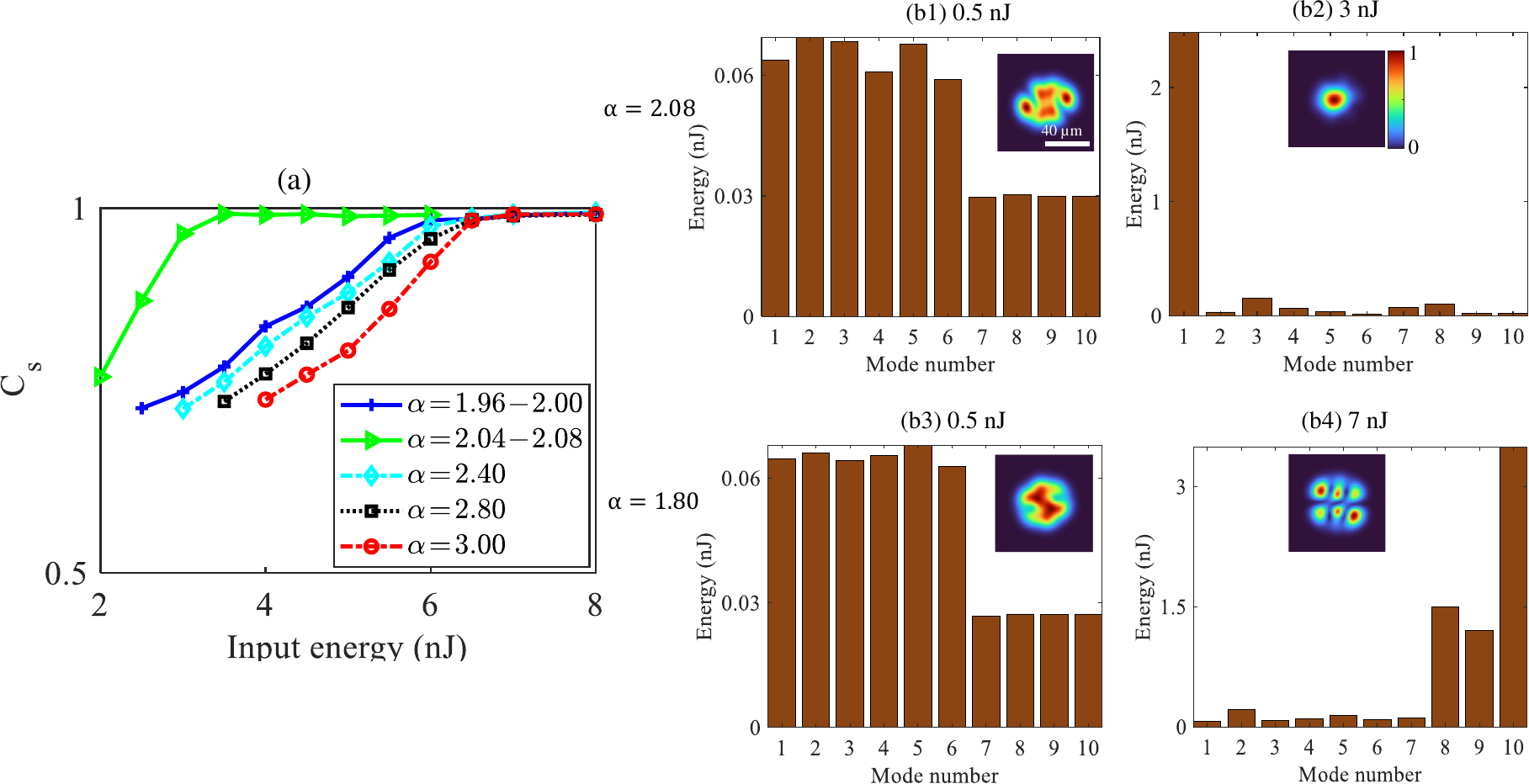}
\centering
\caption{Evolution of MM beam dynamics with increasing input energy for different refractive index distributions characterized by the index exponent $\alpha$. (a) Variation of the spatial intensity correlation parameter $C_s$ as a function of input energy, illustrating the transition from a speckled MM profile to a condensed quasi-Gaussian beam. (b1)–(b4) Output modal energy distribution across guided modes at selected input energies, along with corresponding output intensity profiles (insets), for representative values of $\alpha$. At low input energies, energy is distributed across multiple modes, while with increasing input energy, energy progressively concentrates into LOMs. Efficient condensation into the fundamental mode is observed for $\alpha \approx 2.04$--$2.08$, whereas for lower values of $\alpha$ (e.g., $\alpha = 1.80$), the energy redistribution show signatures of reverse energy flow toward HOMs.}
\label{fig6}
\end{figure*}

\noindent
A wide range of intriguing nonlinear phenomena have been observed in MMFs in recent years. Among these, one of the most remarkable is the spatial condensation of a MM beam into a well-defined, bell-shaped, quasi-Gaussian profile with increasing input energy during propagation through GRIN-MMFs. Such spatial beam self-organization can occur in two fundamentally distinct physical regimes.

In the normal dispersion regime ($\beta_{2} > 0$), the spatial reorganization of a MM optical field arises primarily from IM-FWM, which redistributes energy among the guided modes. This process, commonly referred to as \textit{Kerr beam self-cleaning}, leads to a progressive concentration of energy into LOMs, resulting in an improved beam profile \cite{krupa2017spatial}. From a statistical perspective, this evolution can be interpreted as a tendency of a MM system towards a Rayleigh--Jeans type thermal equilibrium, where the modal energy distribution is described by the thermalization of a classical nonlinear wave \cite{ferraro2023spatial, pourbeyram2022direct}. In contrast, in the anomalous dispersion regime ($\beta_{2} < 0$) relevant to MM soliton formation, spatial condensation is governed by the combined action of Kerr and Raman nonlinearities. In this case, energy transfer towards the fundamental mode is predominantly driven by intermodal Raman scattering, and the resulting modal distribution can be described by a weighted Bose-–Einstein like statistics \cite{zitelli2025new}. Although both mechanisms lead to improved beam quality and a bell-shaped output profile, they are governed by fundamentally different nonlinear processes.

To elucidate the role of refractive index distribution in the spatial condensation of a MM soliton beam, we numerically investigate the evolution of the output spatial intensity profile with increasing input energy for different values of the index exponent $\alpha$. The results are presented in Fig. \ref{fig5}. At low input energies ($E_{in}=0.5$\thinspace{nJ}), the output intensity exhibits a speckled structure, resulting from the incoherent superposition of multiple spatial modes propagating with different phase velocities. In this regime, nonlinear interactions are negligible, and the output modal energy distribution remains close to the initial excitation conditions, as shown in Figs. \ref{fig6}(b1) and \ref{fig6}(b3).

With further increase in input energy, the system undergoes a transition into the MM soliton regime, characterized by an abrupt and efficient transfer of energy towards the fundamental mode. This transition is driven by the combined action of intermodal Raman scattering and IM-FWM, which together enhance nonlinear mode coupling and promote irreversible energy flow toward LOMs. As a result, the output beam evolves into a well-defined, quasi-Gaussian profile with significantly improved spatial quality signifying the condensation of the MM soliton into a fundamental mode Raman soliton, and has been shown in Fig. \ref{fig6}(b2).

Importantly, the threshold energy required for this condensation process is strongly influenced by the core refractive index distribution. As observed from Fig. \ref{fig5}, the condensation threshold increases as the index exponent $\alpha$ deviates from the optimal range $\alpha \approx 2.04\text{--}2.08$. Within this optimal regime, the refractive index profile closely approximates an ideal parabolic distribution, leading to maximal intermodal overlap and enhanced nonlinear coupling. Consequently, both IM-FWM and intermodal Raman scattering are significantly strengthened, enabling efficient energy transfer and MM soliton condensation at lower input energies.

In contrast, as the index exponent $\alpha$ deviates from this optimal range, the refractive index distribution departs from the ideal parabolic form, leading to increased modal dispersion and enhanced modal walk-off. The resulting temporal separation among co-propagating modes reduces the effective interaction length over which nonlinear processes can occur. Consequently, the efficiency of intermodal nonlinear interactions is significantly diminished, weakening the energy exchange among modes and increasing the threshold for MM soliton condensation. In extreme cases, this reduction in nonlinear coupling may even inhibit complete condensation, resulting in a residual MM structure in the output beam, as shown in Fig. \ref{fig5}(a1)-Fig. \ref{fig5}(e1) for $\alpha = 1.80$. For $\alpha = 1.80$, the deviation from the parabolic profile alters the modal overlap landscape and enhances modal walk-off, leading to a redistribution of nonlinear coupling pathways. As a consequence, the direction of energy flow is modified, and a net transfer of energy towards HOMs is observed, in contrast to the conventional condensation of MM soliton beams towards the fundamental mode [Fig. \ref{fig6}(b4)]. This observation highlights that the refractive index distribution not only controls the strength of nonlinear interactions but can also fundamentally alter the directionality of energy transfer in MM soliton systems.

Interestingly, this behavior bears a strong qualitative resemblance to the reverse energy flow reported in the normal dispersion regime, where it has been interpreted in terms of Rayleigh–Jeans thermalization toward negative-temperature equilibrium states \cite{baudin2023observation}. Within such a framework, the system evolves toward an inverted modal population characterized by preferential occupation of HOMs. While the present study operates in the anomalous dispersion regime and involves the formation of coherent MM soliton structures rather than weakly nonlinear thermalized states, the observed energy redistribution toward HOMs suggests the emergence of a dynamical analogue of such inverted population states. Additionally, in our case, this behavior originates from the combined effects of enhanced modal walk-off and modified intermodal nonlinear interactions, which suppress effective energy transfer towards the fundamental mode and instead favor redistribution toward HOMs. This highlights that, although the underlying mechanisms differ, similar reverse energy flow can arise in the MM soliton regime, consistent with thermodynamic interpretations that have previously been used to describe MM optical systems.

Figure \ref{fig6}(a) shows the variation of the intensity correlation parameter, $C_s$, as a function of input energy for different values of the index exponent $\alpha$. The parameter $C_s$ serves as a quantitative measure of the degree of intensity correlation between the observed output profile and the fundamental mode profile, where lower values indicate a weak resemblance to the fundamental mode, corresponding to a speckled and highly MM output, whereas higher values signify a strong resemblance, indicating a well-condensed, quasi-Gaussian beam dominated by the fundamental mode.

At low input energies, $C_s$ assumes relatively small values for all $\alpha$, reflecting the weak resemblance of the output intensity profile with the fundamental mode due to the incoherent superposition of multiple spatial modes and the absence of intermodal energy exchanges. As the input energy increases, $C_s$ increases monotonically, indicating the onset of nonlinear intermodal coupling and the resulting progressive spatial reorganization of the MM beam toward a more coherent, fundamental-mode like profile.

A strong dependence of $C_s$ on the index exponent $\alpha$ is observed. For values of $\alpha$ within or close to the optimal range (e.g., $\alpha \approx 2.04$--$2.08$), the correlation parameter increases rapidly with input energy and approaches unity at relatively lower energies. This behavior signifies efficient intermodal nonlinear interactions at relatively lower input energies, leading to rapid condensation of the MM beam into a quasi-Gaussian profile.

In contrast, for values of $\alpha$ away from the optimal range ($\alpha = 2.40, 2.80, 3.00$), the increase in $C_s$ is more gradual, and higher input energies are required to achieve comparable levels of beam condensation. This behavior indicates reduced efficiency of intermodal nonlinear coupling, primarily due to enhanced modal walk-off, which limits the effective interaction length. Nevertheless, at sufficiently high input energies, all cases approach $C_s \approx 1$, indicating that strong nonlinear effects ultimately drive the system toward a condensed state.

\section{Conclusions}
\noindent
In summary, we have carried out a comprehensive numerical investigation of the influence of the core refractive index distribution, characterized by the index exponent $\alpha$, on the evolution and characteristic properties of MM solitons in GRIN-MMFs. In particular, we identify an optimal range of $\alpha \approx 2.04\text{--}2.08$, within which MM solitons with minimum pulsewidth and lower characteristic energy are formed due to reduced modal walk-off and enhanced nonlinear intermodal interactions. In addition, we show that the refractive index distribution plays a crucial role in governing the efficiency of the MM soliton condensation process. Importantly, our study also reveals that deviation from this optimal range can significantly alter the energy transfer dynamics, including the emergence of reverse energy flow toward HOMs for lower values of $\alpha$, showing resemblance to the thermalization of MM optical fields toward a negative-temperature equilibrium state. Furthermore, we demonstrate that the characteristic Raman-induced spectral redshift of MM solitons can be effectively controlled by tailoring the refractive index distribution. 
Overall, our findings establish the refractive index distribution as key parameters for controlling MM soliton dynamics. These insights may pave the way for improved design and optimization of MM fiber systems for applications requiring tailored spatiotemporal soliton properties, including high-power fiber lasers and spatial-division multiplexed communication systems.

\section*{Acknowledgement}
\noindent
We acknowledge the funding support from the Science and Engineering Research Board (SERB), India (Grant no. CRG/2021/003060). Love Kumar Sharma acknowledges the financial support by IIT Ropar.

%\bibliography{Article_files/Bibliography}

\end{document}